# BELL'S THEOREM : THE NAIVE VIEW OF AN EXPERIMENTALIST[†]

Alain Aspect

Institut d'Optique Théorique et Appliquée
Bâtiment 503-Centre universitaire d'Orsay
91403 ORSAY Cedex – France
alain.aspect@iota.u-psud.fr

## 1. INTRODUCTION

It is a real emotion to participate to this conference in commemoration of John Bell. I first met him in 1975, a few months after reading his famous paper[1]. I had been so strongly impressed by this paper, that I had immediately decided to do my « thèse d'état » – which at that time, in France, could be a really long work – on this fascinating problem. I definitely wanted to carry out an experiment « in which the settings are changed during the flight of the particles », as suggested in the paper, and I had convinced a young professor of the Institut d'Optique, Christian Imbert, to support my project and to act as my thesis advisor. But he had advised me to first go to Geneva, and to discuss my proposal with John Bell. I got an appointment without delay, and I showed up in John's office at CERN, very impressed. While I was explaining my planned experiment, he silently listened. Eventually, I stopped talking, and the first question came: "Have you a permanent position?" After my positive answer, he started talking of physics, and he definitely encouraged me, making it clear that he would consider the implementation of variable analysers a fundamental improvement. Beyond his celebrated sense of humour, his answer reminds me of the general atmosphere at that time about raising questions on the foundations of quantum mechanics. Quite frequently it was open hostility, and in the best case, it would provoke an ironical reaction: "Quantum Mechanics has been vindicated by such a large amount of work by the smartest theorists and experimentalists, how can you hope to find anything with such a simple scheme, in optics, a science of the XIX[th] century?" In addition to starting the experiment, I had then to develop a line of argument to try to convince the physicists I met (and among them some had to give their opinion about funding my project). After some not so successfull tentatives of quite sophisticated pleas, I eventually found out that it was much more efficient to explain the very simple and naive way in which I had understood Bell's theorem. And to my great surprise, that simple presentation was very convincing even with the most theoretically inclined interlocutors. I was lucky enough to be able to present it in front of John Bell himself, and he apparently appreciated it. I am therefore going to explain now how I understood Bell's theorem twenty five years ago, and I hope to be able to communicate the shock I received, that was so strong that I spent eight years of my life working on this problem.

---

[†] This text was prepared for a talk at a conference in memory of John Bell, held in Vienna in December 2000. It has been published in "Quantum [Un]speakables – From Bell to Quantum information", edited by R. A. Bertlmann and A. Zeilinger, Springer (2002).

This written transcription of my presentation is partly based on a paper that was published two decades ago as a proceedings of a conference, not so easy to find nowadays[2]. The first part of the paper aims at explaining what are Bell's theorem and Bell's inequalities, and why I find it so important. It is followed by a rapid review of the *first generation of experimental tests of Bell's inequalities with pairs of entangled photons*, carried out between 1971 and 1976. I am glad that most of the heroes of this seeding work are present at this meeting. I give then a more detailed description of the three *experiments of second generation*, that we performed at the Institut d'Optique d'Orsay between 1976 and 1982, with a dramatically improved source of pairs of entangled photons, using a non linear two photon laser excitation of atomic radiative cascades. The last part gives an overview of the *experiments of third generation*, developed since the late 80's, and carried out with pairs of entangled photons produced in non linear parametric down conversion: these experiments can close most of the loopholes still left open in the second generation experiments. I am deliberately concentrating on optics experiments, since they are at the present time the most convincing and the closest to the ideal GedankenExperiment, but the interested reader should be aware that other systems do exist, in other domains of physics, that may offer the possibility to perform as convincing experiments.

In the first part of this presentation (sections 2 to 6), we will see that Bell's Inequalities provide a quantitative criterion to test « reasonable » Supplementary Parameters Theories versus Quantum Mechanics. Following Bell, I will first explain the motivations for considering supplementary parameters theories: the argument is based on an analysis of the famous Einstein-Podolsky-Rosen (EPR) Gedankenexperiment[3]. Introducing a reasonable Locality Condition, we will then derive Bell's theorem, which states:

i. that Local Supplementary Parameters Theories are constrained by Bell's Inequalities;

ii. that certain predictions of Quantum Mechanics violate Bell's Inequalities, and therefore that Quantum Mechanics is incompatible with Local Supplementary Parameters Theories.

We will then point out that a fundamental assumption for this conflict is the *Locality assumption*. And we will show that in a more sophisticated version of the E.P.R. thought experiment (« timing experiment »), the Locality Condition may be considered a *consequence of Einstein's Causality*, preventing faster-than-light interactions.

The purpose of this first part is to convince the reader that the *formalism leading to Bell's Inequalities is very general and reasonable*. What is *surprising* is that such a reasonable formalism *conflicts with Quantum Mechanics*. In fact, situations exhibiting a *conflict are very rare*, and Quantum Optics is the domain where the most significant tests of this conflict have been carried out (sections 7 to 11).

## 2. WHY SUPPLEMENTARY PARAMETERS ? THE EINSTEIN-PODOLSKY-ROSEN-BOHM GEDANKENEXPERIMENT

### *2.1. Experimental scheme*

Let us consider the optical variant of the Bohm's version[4] of the E.P.R. Gedankenexperiment (Fig. 1). A source *S* emits a pair of photons with different frequencies $\nu_1$ and $\nu_2$, counterpropagating along *Oz*. Suppose that the polarization part of the state vector describing the pair is:

$$|\Psi(\nu_1,\nu_2)\rangle = \frac{1}{\sqrt{2}}\{|x,x\rangle + |y,y\rangle\} \quad (1)$$

where $|x\rangle$ and $|y\rangle$ are linear polarizations states. This state is remarkable : it cannot be factorized into a product of two states associated to each photon, so we cannot ascribe any well defined state to each photon. In particular, we cannot assign any polarization to each photon. Such a state describing a system of several objects that can only be thought of globally, is an *entangled state*.

We perform linear polarization measurements on the two photons, with analysers *I* and *II*. The analyser *I*, in orientation **a**, is followed by two detectors, giving results + or −, corresponding to a linear polarization found parallel or perpendicular to **a**. The analyser *II*, in orientation **b**, acts similarly‡.

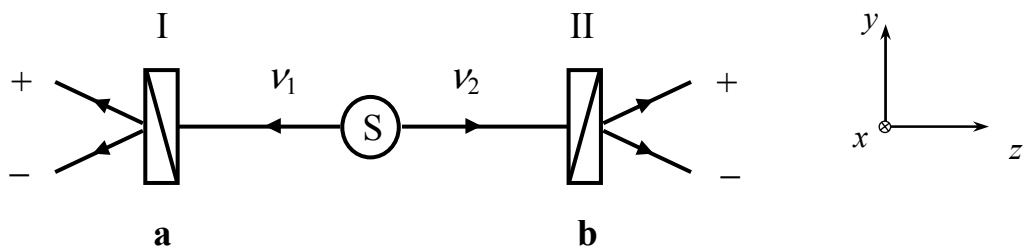

*Figure 1. Einstein-Podolsky-Rosen-Bohm Gedankenexperiment with photons. The two photons $\nu_1$ and $\nu_2$, emitted in the state $|\Psi(1,2)\rangle$ of Equation (1), are analyzed by linear polarizers in orientations **a** and **b**. One can measure the probabilities of single or joint detections in the output channels of the polarizers.*

It is easy to derive the Quantum Mechanical predictions for these measurements of polarization, single or in coincidence. Consider first the singles probabilities $P_\pm(\mathbf{a})$ of getting the results ± for the photon $\nu_1$, and similarly, the singles probabilities $P_\pm(\mathbf{b})$ of obtaining the results ± on photon $\nu_2$. Quantum Mechanics predicts:

---

‡ There is a one-to-one correspondance with the EPR Bohm Gedankenexperiment dealing with a pair of spin 1/2 particles, in a singlet state, analysed by two orientable Stern-Gerlach filters.

$$P_+(\mathbf{a}) = P_-(\mathbf{a}) = 1/2$$
$$P_+(\mathbf{b}) = P_-(\mathbf{b}) = 1/2 \qquad \text{(Q.M.)} \qquad (2)$$

These results are in agreement with the remark that we cannot assign any polarization to each photon, so that each individual polarization measurement gives a random result.

Let us now consider the probabilities $P_{\pm\pm}(\mathbf{a},\mathbf{b})$ of joint detections of $v_1$ and $v_2$ in the channels + or − of polarisers *I* or *II*, in orientations **a** and **b**. Quantum mechanics predicts :

$$P_{++}(\mathbf{a},\mathbf{b}) = P_{--}(\mathbf{a},\mathbf{b}) = \frac{1}{2}\cos^2(\mathbf{a},\mathbf{b})$$
$$P_{+-}(\mathbf{a},\mathbf{b}) = P_{-+}(\mathbf{a},\mathbf{b}) = \frac{1}{2}\sin^2(\mathbf{a},\mathbf{b}) \qquad \text{(Q.M.)} \qquad (3)$$

We are going to show that these quantum mechanical predictions have far reaching consequences.

## *2.2. Correlations*

Consider first the particular situation $(\mathbf{a},\mathbf{b}) = 0$, where polarisers are parallel. The Quantum Mechanical predictions for the the joint detection probabilities (equations 3) are :

$$P_{++}(\mathbf{a},\mathbf{a}) = P_{--}(\mathbf{a},\mathbf{a}) = \frac{1}{2}$$
$$P_{+-}(\mathbf{a},\mathbf{a}) = P_{-+}(\mathbf{a},\mathbf{a}) = 0 \qquad (4)$$

According to this result, and taking into account (2), we conclude that when the photon $v_1$ is found in the + channel of polarizer *I*, $v_2$ is found *with certainty* in the + channel of *II* (and similarly for the − channels). For parallel polarizers, there is thus a *total correlation* between the individually random results of measurements of polarization on the two photons $v_1$ and $v_2$.

A convenient way to measure the amount of correlations between random quantities, is to calculate the correlation coefficient. For the polarization measurements considered above, it is equal to

$$E(\mathbf{a},\mathbf{b}) = P_{++}(\mathbf{a},\mathbf{b}) + P_{--}(\mathbf{a},\mathbf{b}) - P_{+-}(\mathbf{a},\mathbf{b}) - P_{+-}(\mathbf{a},\mathbf{b}) \qquad (5)$$

Using the prediction (3) of Quantum Mechanics, we find a correlation coefficient

$$E_{QM}(\mathbf{a},\mathbf{b}) = \cos 2(\mathbf{a},\mathbf{b}) \qquad (6)$$

In the particular case of parallel polarizers ( $(\mathbf{a},\mathbf{b}) = 0$ ), we find $E_{QM}(0) = 1$ : this confirms that the correlation is total.

In conclusion, the quantum mechanical calculations suggest that although each individual measurement gives random results, these random results are correlated, as expressed by equation (6). For parallel (or perpendicular) orientations of the polarizers, the correlation is total ($|E_{QM}| = 1$).

## 2.3. Difficulty of an image derived from the formalism of Quantum Mechanics

As a naive physicist, I like to raise the question of finding a simple image to understand these strong correlations. The most natural way to find an image may seem to follow the quantum mechanical calculations leading to (3). In fact, there are several ways to do this calculation. A very direct one is to project the state vector (1) onto the eigenvector corresponding to the relevant result. This gives immediately the joint probabilities (3). However, since this calculation bears on state vectors describing globally the two photons, I do not know how to build a picture in our ordinary space.

In order to overcome this problem, and to identify separately the two measurements happening on both ends of the experiment, we can split the joint measurement in two steps. Suppose for instance that the measurement on photon $\nu_1$ takes place first, and gives the result + , with the polarizer I in orientation **a**. The + result (associated with the polarization state $|\mathbf{a}\rangle$) has a probability of $1/2$. To proceed with the calculation, we must then use the postulate of reduction of the state vector, which states that after this measurement, the new state vector $|\Psi'(\nu_1, \nu_2)\rangle$ describing the pair is obtained by projection of the initial state vector $|\Psi(\nu_1, \nu_2)\rangle$ (equation 1) onto the eigenspace associated to the result + : this two dimensional eigenspace has a basis $\{|\mathbf{a}, x\rangle, |\mathbf{a}, y\rangle\}$. Using the corresponding projector, we find after a little algebra

$$|\Psi'(\nu_1, \nu_2)\rangle = |\mathbf{a}, \mathbf{a}\rangle \tag{7}$$

This means that immediately after the first measurement, photon $\nu_1$ takes the polarization $|\mathbf{a}\rangle$ : this is obvious because it has been measured with a polarizer oriented along **a**, and the result + has been found. More surprisingly, the distant photon $\nu_2$, which has not yet interacted with any polarizer, has also been projected into the state $|\mathbf{a}\rangle$ with a well defined polarization, parallel to the one found for photon $\nu_1$. This surprising conclusion however leads to the correct final result (3), since a straightforward application of Malus law shows that a subsequent measurement performed along **b** on photon $\nu_2$ will lead to

$$P_{++}(\mathbf{a}, \mathbf{b}) = \frac{1}{2} \cos^2(\mathbf{a}, \mathbf{b}) \tag{8}$$

The calculation in two steps therefore gives the same result as the direct calculation. But in addition it suggests a picture for the two steps measurement:

i. Photon $v_1$, which had not a well defined polarization before its measurement, takes the polarization associated to the obtained result, at the moment of its measurement: this is not surprising.

ii. When the measurement on $v_1$ is done, photon $v_2$, which had not a well defined polarization before this measurement, is projected into a state of polarization parallel to the result of the measurement on $v_1$. This is very surprising, because this change in the description of $v_2$ happens instantaneously, whatever the distance between $v_1$ and $v_2$ at the moment of the first measurement.

This picture seems in contradiction with relativity. According to Einstein, what happens in a given region of space-time cannot be influenced by an event happening in a region of space-time that is separated by a space like interval. It therefore not unreasonable to try to find more acceptable pictures for « understanding » the EPR correlations. It is such a picture that we consider now.

## *2.4. Supplementary parameters*

Correlations between distant measurements on two separated systems that had previously interacted are common in the classical world. For instance, if a mechanical object with a null linear (or angular) momentum is split in two parts by some internal repulsion, the linear (or angular) momenta of the two separated parts remain equal and opposite in the case of a free evolution. In the general case where each fragment is submitted to some interaction, the two momenta remain correlated since they are at each moment determined by their initial values, which had a perfectly defined sum.

It is tempting to use such a classical picture to render an account of the EPR correlations, in term of common properties of the two systems. Let us consider again the perfect correlation of polarization measurements in the case of parallel polarisers $(\mathbf{a},\mathbf{b}) = 0$. When we find + for $v_1$, we are sure to find also + for $v_2$. We are thus led to admit that there is some property (Einstein said « an element of physical reality ») pertaining to this particular pair, and determining the result ++. For another pair, when the results is − −, we can similarly invoke a common property, determining the result − −. It is then sufficient to admit that half the pairs are emitted with the property ++, and half with the property − −, to reproduce all the results of measurement in this configuration. Note however that such properties, differing from one pair to another one, are not taken into account by the Quantum Mechanical state vector $|\Psi(v_1, v_2)\rangle$ which is the same for all pairs. This is why we can conclude with Einstein that *Quantum Mechanics is not complete*. And this is why such additional properties are referred to as « *supplementary parameters* », or « *hidden-variables* »[*]

---

[*] Einstein actually did not speak of « hidden variables » or « supplementary parameters », but rather of « elements of the physical reality ». Accordingly, many authors refer to « realistic theories » rather than to « hidden variable theories », or to « supplementary variable theories ».

As a conclusion, it seems possible to « understand » the EPR correlations by such a classical-looking picture, involving supplementary parameters differing from one pair to another one. It can be hoped to recover the statistical Quantum Mechanical predictions when averaging over the supplementary parameters. It seems that so was Einstein's position[5,6,7]. Note that at this stage of the reasoning, a commitment to this position does not contradict quantum mechanics: there is no logical problem to fully accept the predictions of quantum mechanics, *and* to invoke supplementary parameters giving an acceptable picture of the EPR correlations. It amounts to considering Quantum Mechanics as the Statistical Mechanics description of a deeper level.

## 3. BELL'S INEQUALITIES

### 3.1. Formalism

Three decades after the EPR paper, Bell translated into mathematics the consequences of the preceding discussion, and he explicitly introduced supplementary parameters, denoted $\lambda$. Their distribution on an ensemble of emitted pairs is specified by a probability distribution $\rho(\lambda)$, such that

$$\rho(\lambda) \geq 0$$
$$\int d\lambda \rho(\lambda) = 1 \tag{9}$$

For a given pair, characterized by a given supplementary parameter $\lambda$, the results of measurements are given by the bivalued functions

$$A(\lambda, \mathbf{a}) = \pm 1 \quad \text{at analyzer I (in orientation } \mathbf{a}\text{)}$$
$$B(\lambda, \mathbf{b}) = \pm 1 \quad \text{at analyzer II (in orientation } \mathbf{b}\text{)} \tag{10}$$

A particular Supplementary Parameter Theory is completely defined by the explicit form of the function $\rho(\lambda)$, $A(\lambda,\mathbf{a})$ and $B(\lambda,\mathbf{b})$. It is then easy to express the probabilities of the various results of measurements. For instance, noting that the function $\frac{1}{2}[A(\lambda, \mathbf{a}) + 1]$ assumes the value +1 for the + result, and 0 otherwise (and similarly $\frac{1}{2}[1 - B(\lambda, \mathbf{b})]$ assumes the value +1 for the − result, and 0 otherwise, we can write

$$P_+(\mathbf{a}) = \int d\lambda \rho(\lambda) \frac{[A(\lambda, \mathbf{a}) + 1]}{2}$$
$$P_{+-}(\mathbf{a}, \mathbf{b}) = \int d\lambda \rho(\lambda) \frac{[A(\lambda, \mathbf{a}) + 1]}{2} \frac{[1 - B(\lambda, \mathbf{b})]}{2} \tag{11}$$

Similarly, the correlation function assumes the simple form

$$E(\mathbf{a}, \mathbf{b}) = \int d\lambda \rho(\lambda) A(\lambda, \mathbf{a}) B(\lambda, \mathbf{b}) \tag{12}$$

*3.2. A (naive) example of supplementary parameters theory*

As an example of Supplementary Parameter Theory, we present a model where each photon travelling along Oz is supposed to have a well defined linear polarization, determined by its angle $(\lambda_1 \text{ or } \lambda_2)$ with the *x* axis. In order to account for the strong correlation, we assume that the two photons of a same pair are emitted with the same linear polarization, defined by the common angle $\lambda$ (figure 2).

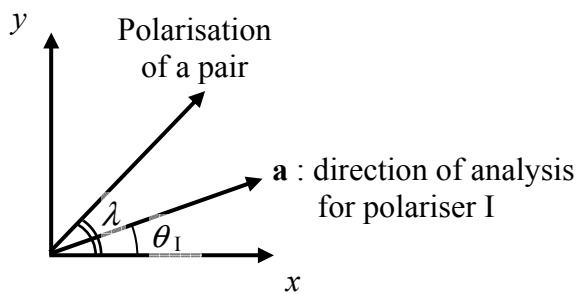

*Figure 2 - **The naive example**. Each pair of photons has a « direction of polarisation », defined by $\lambda$, which is the supplementary parameter of the model. Polariser I makes a polarisation measurement along **a**, at an angle $\theta_I$ from the x axis.*

The polarisation of the various pairs is randomly distributed, according to the probability distribution $\rho(\lambda)$ that we take rotationally invariant:

$$\rho(\lambda) = \frac{1}{2\pi} \tag{13}$$

To complete our model, we must give an explicit form for the functions $A(\lambda,\mathbf{a})$ and $B(\lambda,\mathbf{b})$. We take the following form

$$A(\lambda,\mathbf{a}) = sign\{\cos 2(\theta_I - \lambda)\}$$
$$B(\lambda,\mathbf{b}) = sign\{\cos 2(\theta_{II} - \lambda)\} \tag{14}$$

where the angles $\theta_I$ and $\theta_{II}$ indicate the orientations of the polarisers. Note that these forms are very reasonable: $A(\lambda,\mathbf{a})$ assumes the value +1 when the polarisation of photon $\nu_1$ makes an angle less than $\frac{\pi}{4}$ with the direction of analysis **a**, and $-1$ for the complementary case (polarisation closer to the perpendicular to **a**).

With this explicit model, we can use equations (11) to calculate the probabilities of the various measurements. We find for instance single probabilities

$$P_+(\mathbf{a}) = P_-(\mathbf{a}) = P_+(\mathbf{b}) = P_-(\mathbf{b}) = \frac{1}{2} \tag{15}$$

identical to the Quantum Mechanical results. The model also allows us to calculate the joint probabilities, or equivalently the correlation function, and we find, using (12) :

$$E(\mathbf{a},\mathbf{b}) = 1 - 4\frac{|\theta_I - \theta_{II}|}{\pi} = 1 - 4\frac{|(\mathbf{a},\mathbf{b})|}{\pi} \tag{16}$$

for $\quad -\frac{\pi}{2} \leq \theta_I - \theta_{II} \leq \frac{\pi}{2}$

This is a remarkable result. Note first that E(**a,b**) depends only on the relative angle (**a,b**), as the Quantum Mechanical prediction (6). Moreover, as shown on figure 3, the difference beween the predictions of the simple supplementary parameters model and the quantum mechanical predictions is always small, and the agreement is exact for the angles 0 and π€/€2, *i.e.* cases of total correlation. This result, obtained with an extremely simple supplementary parameters model, is very encouraging, and it might be hoped that a more sophisticated model could be able to reproduce exactly the Quantum Mechanical predictions. *Bell's discovery is the fact that the search for such models is hopeless*, as we are going to show now.

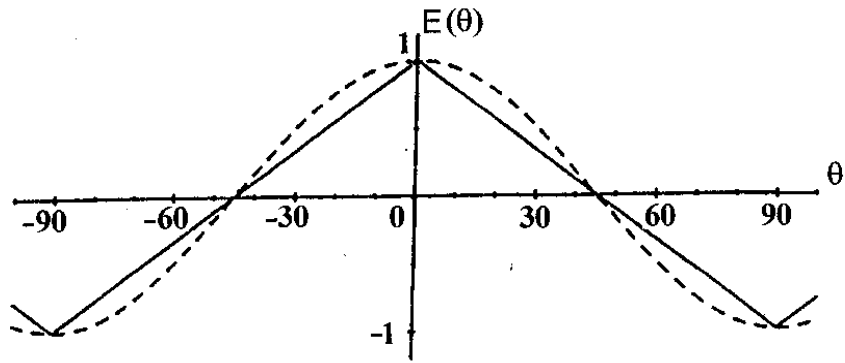

**Figure 3** - *Polarisation correlation coefficient, as a function of the relative orientation of the polarisers: (i) Dotted line : Quantum Mechanical prediction ; (ii) solid line : the naive model.*

## 3.3. Bell's Inequalities

There are many different forms, and demonstrations of Bell'inequalities. We give here a very simple demonstration leading to a form directly applicable to the experiments**.

Let us consider the quantity

$$s = A(\lambda,\mathbf{a}).B(\lambda,\mathbf{b}) - A(\lambda,\mathbf{a}).B(\lambda,\mathbf{b'}) + A(\lambda,\mathbf{a'}).B(\lambda,\mathbf{b}) + A(\lambda,\mathbf{a'}).B(\lambda,\mathbf{b'})$$
$$= A(\lambda,\mathbf{a})[B(\lambda,\mathbf{b}) - B(\lambda,\mathbf{b'})] + A(\lambda,\mathbf{a'})[B(\lambda,\mathbf{b}) + B(\lambda,\mathbf{b'})] \quad (17)$$

Remembering that the four numbers $A$ and $B$ take only the values $\pm 1$, a simple inspection of the second line of (17) shows that

$$s(\lambda,\mathbf{a},\mathbf{a'},\mathbf{b},\mathbf{b'}) = \pm 2 \quad (18)$$

The average of $s$ over $\lambda$ is therefore comprised between $+2$ and $-2$

$$-2 \leq \int d\lambda \rho(\lambda).s(\lambda,\mathbf{a},\mathbf{a'},\mathbf{b},\mathbf{b'}) \leq 2 \quad (19)$$

According to (12), we can rewrite these inequalities

$$-2 \leq S(\mathbf{a},\mathbf{a'},\mathbf{b},\mathbf{b'}) \leq 2 \quad (20)$$

with

$$S(\mathbf{a},\mathbf{a'},\mathbf{b},\mathbf{b'}) = E(\mathbf{a},\mathbf{b}) - E(\mathbf{a},\mathbf{b'}) + E(\mathbf{a'},\mathbf{b}) + E(\mathbf{a'},\mathbf{b'}) \quad (21)$$

These are B.C.H.S.H. inequalities, *i.e.* Bell's inequalitites as generalized by Clauser, Horne, Shimony, Holt[8]. They bear upon the combination $S$ of the four polarization correlation coefficients, associated to two directions of analysis for each polarizer (**a** and **a'** for polarizer *I*, **b** and **b'** for polarizer *II*). Note that they apply to any Supplementary Parameter Theory of the very general form defined in section 3.1 (equations 9, 10, and 12), of which our naive model is only an example.

---

** It is important to distinguish between inequalities which show a mathematical contradiction with quantum mechanics, but without the possibility of an experimental test with (necessarily) imperfect apparatus, and inequalities allowing an experimental test provided that the experimental imperfections remain in certain limits.

## 4. CONFLICT WITH QUANTUM MECHANICS

*4.1. Evidence*

We can use the predictions (6) of Quantum Mechanics for EPR pairs, to evaluate the quantity $S(\mathbf{a},\mathbf{a'},\mathbf{b},\mathbf{b'})$ defined by equation (21). For the particular set of orientations shown on Figure 4.a, the result is

$$S_{QM} = 2\sqrt{2} \qquad (22)$$

This quantum mechanical prediction definitely conflicts with the Bell's inequality (20) which is valid for any Supplementary Parameter Theory of the general form defined in §3.1.

We have thus found a situation where the quantum mechanical predictions cannot be reproduced (mimicked) by Supplementary Parameters Theories. This is the essence of Bell's theorem: it is impossible to find a Supplementary Parameter Theory, of the general form defined in § 3.1, that reproduces **all** the predictions of quantum mechanics. This statement is the generalisation of what appears on Figure 3, for the particular supplementary parameter model considered in § 3.2: the model exactly reproduces the predictions of quantum mechanics for some particular angles (0, π/4, π/2), but it somewhat deviates at other angles. The importance of Bell's theorem is that it is not restricted to a particular supplementary parameters model, but it is general.

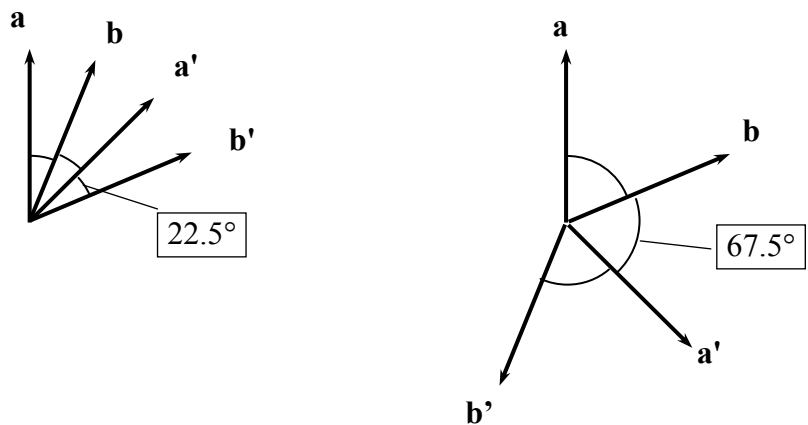

*Figure 4 - Orientations yielding the largest conflict between Bell's inequalities and Quantum Mechanics.*

*4.2. Maximum conflict*

It is interesting to look for the maximum violation of Bell's inequalities by the quantum mechanical predictions. Let us take the quantum mechanical value of $S$:

$$S_{QM}(\mathbf{a},\mathbf{b},\mathbf{a'},\mathbf{b'}) = \cos(\mathbf{a},\mathbf{b}) - \cos(\mathbf{a},\mathbf{b'}) + \cos(\mathbf{a'},\mathbf{b}) + \cos(\mathbf{a'},\mathbf{b'}) \qquad (23)$$

It is a function of the three independant variables $(\mathbf{a},\mathbf{b})$, $(\mathbf{b},\mathbf{a'})$, and $(\mathbf{a'},\mathbf{b'})$. Note that

$$(\mathbf{a}, \mathbf{b'}) = (\mathbf{a}, \mathbf{b}) + (\mathbf{b}, \mathbf{a'}) + (\mathbf{a'}, \mathbf{b'})$$

In order to find the extremum values of $S_{QM}$, we write that the three partial derivatives are null, and we find

$$(\mathbf{a}, \mathbf{b}) = (\mathbf{b}, \mathbf{a'}) = (\mathbf{a'}, \mathbf{b'}) = \theta \tag{24}$$

and

$$\sin \theta = \sin 3\theta \tag{25}$$

We have plotted on Figure 5 the function $S_{QM}(\theta)$ evaluated in the case of condition (24). It shows that the absolute maximum and minimum of $S_{QM}$ are

$$S_{MQ} = 2\sqrt{2} \quad \text{for} \quad \theta = \pm \frac{\pi}{8} \tag{26}$$

$$S_{MQ} = -2\sqrt{2} \quad \text{for} \quad \theta = \pm \frac{3\pi}{8} \tag{27}$$

These values are solutions of (25). The corresponding sets of orientations are displayed on Figures 4. They give the maximum violations of Bell's inequalities.

More generally, Figure 5 shows that there is a full range of orientations leading to a conflict with Bells inequalities. However, it is also clear that there are many sets of orientations for which there is no conflict.

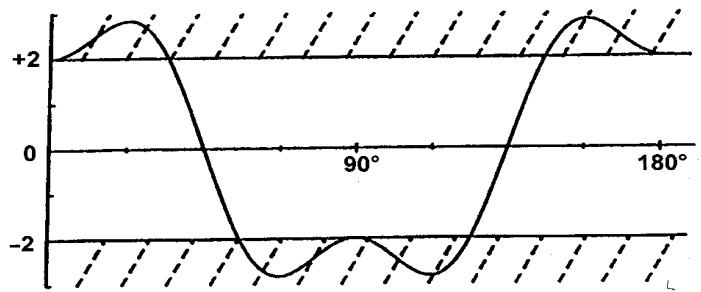

*Figure 5* - *S($\theta$) as predicted by Quantum Mechanics for EPR pairs. The conflict with Bell's inequalities happens when |S| is larger than 2, and it is maximum for the sets of orientations of Figure 4.*

## 5. DISCUSSION : THE LOCALITY CONDITION

We have now established Bell's theorem: Quantum Mechanics conflicts with any Supplementary Parameter Theory as defined in § 3.1, since it violates a consequence (Bell's inequalities) of any such theory. Clearly, it is interesting at this stage to look for the hypotheses underlying the formalism introduced in § 3.1. One may then hope to point out a specific hypothesis responsible for the conflict. We therefore examine now the various hypotheses underlying the Supplementary Parameter Theories introduced in section 3.1.

A first hypothesis is the existence of supplementary parameters. As we have seen, they have been introduced in order to render an account of the correlations at a distance. This hypothesis is strongly linked to a conception of the world, as expressed by Einstein, where the notion of separated physical realities for separated particles is meaningful. It is even possible to derive the existence of supplementary parameters from general statements about the physical reality, in the spirit of Einstein's ideas[9]. An hypothesis in this spirit seems absolutely necessary to obtain inequalities conflicting with quantum mechanics.

The second considered hypothesis is determinism. As a matter of fact, the formalism of section 3.1 is deterministic: once $\lambda$ is fixed, the results $A(\lambda,\mathbf{a})$ and $B(\lambda,\mathbf{b})$ of the polarization measurements are certain. One has speculated that it may be a good reason for a conflict with the non-deterministic formalism of quantum mechanics. In fact, as first shown by Bell[10], and subsequently developped[11], it is easy to generalize the formalism of section 3.1 to *Stochastic* Supplementary Parameter Theories where the deterministic measurement functions $A(\lambda,\mathbf{a})$ and $B(\lambda,\mathbf{b'})$ are replaced by probabilistic functions. One then finds that the Bell's inequalities still hold, and that the conflict does not disappear. It is therefore generally accepted that the deterministic character of the formalism is not the reason for the conflict[12].

The most important hypothesis, stressed by Bell in all his papers, is the local character of the formalism of section 3.1. We have indeed implicitly assumed that the result $A(\lambda,\mathbf{a})$ of the measurement at polarizer *I*, does not depend on the orientation $\mathbf{b}$ of the remote polarizer *II*, and vice-versa. Similarly, it is assumed that the probability distribution $\rho(\lambda)$ (*i.e.* the way in which pairs are emitted) does not depend on the orientations $\mathbf{a}$ and $\mathbf{b}$. This *locality assumption* is crucial: Bell's Inequalities would no longer hold without it. It is indeed clear that the demonstration of § 3.3 fails with quantities such as $A(\lambda,\mathbf{a},\mathbf{b})$ or $\rho(\lambda,\mathbf{a},\mathbf{b})$.

To conclude, there are two hypothesis that seem to be necessary to obtain Bell's inequalities, and consequently a conflict with quantum mechanics :

- distant correlations can be understood by introduction of supplementary parameters carried along by the separated particles, in the spirit of Einstein's ideas that separated objects have separated physical realities.

- the quantities $A(\lambda)$, $B(\lambda)$, and $\rho(\lambda)$ obey the *locality condition*, *i.e.* they do not depend on the orientations of the distant polarizers.

This is why one often claims that Quantum Mechanics conflicts with Local Realism.

# 6. GEDANKENEXPERIMENT WITH VARIABLE ANALYZERS : THE LOCALITY CONDITION AS A CONSEQUENCE OF EINSTEIN'S CAUSALITY

In static experiments, in which the polarizers are held fixed for the whole duration of a run, the Locality Condition must be stated as an assumption. Although highly reasonable, this condition is not prescribed by any fundamental physical law. To quote J. Bell[1] « the settings of the instruments are made sufficiently in advance to allow them to reach some mutual rapport by exchange of signals with velocity less than or equal to that of light ». In that case, the result $A(\lambda)$ of the measurement at polarizer *I* could depend on the orientation **b** of the remote polarizer *II*, and vice-versa. The Locality Condition would no longer hold, nor would Bell's Inequalities.

Bell thus insisted upon the importance of « experiments of the type proposed by Bohm and Aharonov[5], in which the settings are changed during the flight of the particles »[*]. In such a *timing-experiment*, the *locality condition would become a consequence of Einstein's Causality that prevents any faster-than-light influence*.

As shown in our 1975 proposal[13], it is sufficient to switch each polarizer between two particular settings (**a** and **a'** for *I*, **b** and **b'** for *II*). It then becomes possible to test experimentally a large class of Supplementary Parameters Theories: those obeying Einstein's Causality. In such theories, the response of polarizer *I* at time *t*, is allowed to depend on the orientation **b** (or **b'**) of polarizer *II* at times anterior to $t - L/c$ (*L* is the distance between the polarizers). A similar retarded dependence is considered for the probability distribution $\rho(\lambda)$, *i.e.* the way in which pairs are emitted at the source. For random switching times, with both sides uncorrelated, the predictions of these more general « *separable supplementary parameters theories* » are constrained by *generalized Bell's Inequalities*[13], based on Einstein's causality and not on Bell's locality condition.

On the other hand, one can show[14] that the polarization correlations predicted by Quantum Mechanics depend only on the orientations **a** and **b** at the very time of the measurements, and do not involve any retardation terms $L/c$. For a suitable choice of the set of orientations (**a**,**a'**,**b**,**b'**) − for instance the sets displayed on Figure 4 − the Quantum mechanical predictions still conflict with generalized Bell's Inequalities.

In an experiment with time varying polarizers, Bell's theorem therefore states that Quantum Mechanics is incompatible with Supplementary Parameters theories obeying Einstein's causality. Note that Einstein's causality already played an important role in the discussions leading to the notion of supplementary parameters, or equivalently of an independant physical reality for each separated subsystem[6]. It therefore does not seem exagerate to conclude that in a scheme with time varying polarizers, Bell's theorem establishes a *contradiction between Quantum Mechanics and a description of the world in the spirit of Einstein's ideas*. Note however that *Einstein did not know Bell's theorem*, and he could logically think that his world view was compatible with all the algebraic predictions of quantum mechanics. *We do not know what would have been his reaction in front of the contradiction revealed by Bell's theorem*.

---

[*] The idea was already expressed in Bohm's book[4].

# 7. FROM BELL'S THEOREM TO A REALISTIC EXPERIMENT

## *7.1. Experimentallly testing Bell's inequalities*

With Bell's theorem, the debate on the possibility (or necessity) of completing quantum mechanics changed dramatically. It was no longer a matter of philosophical position (realism versus positivism...), or of personal taste. It became possible to settle the question by an experiment. If one can produce pairs of photons (or of spin 1/2 particles) in an EPR state, and measure the 4 coincidence rates $N_{\pm\pm}(\mathbf{a},\mathbf{b})$ with detectors in the output channels of the polarizers (or Stern-Gerlach filters), one obtains the polarization correlation coefficient, for polarizers in orientations **a** and **b** :

$$E(\mathbf{a},\mathbf{b}) = \frac{N_{++}(\mathbf{a},\mathbf{b}) - N_{+-}(\mathbf{a},\mathbf{b}) - N_{-+}(\mathbf{a},\mathbf{b}) + N_{--}(\mathbf{a},\mathbf{b})}{N_{++}(\mathbf{a},\mathbf{b}) + N_{+-}(\mathbf{a},\mathbf{b}) + N_{-+}(\mathbf{a},\mathbf{b}) + N_{--}(\mathbf{a},\mathbf{b})} \quad (28)$$

By performing four measurements of this type in orientations (**a**,**b**), (**a**,**b'**), (**a'**,**b**), and (**a'**,**b'**), one obtains a measured value $S_{\exp}(\mathbf{a},\mathbf{a'},\mathbf{b},\mathbf{b'})$ for the quantity *S* defined in equation (21). Choosing a situation where quantum mechanics predicts that this quantity violates Bell's inequalities (20), we have a test allowing one to discriminate between quantum mechanics and any local supplementary parameter theory. If in addition we use a scheme with variable polarizers, we even test the more general class of « *separable* » (or *causal* in the relativistic sense) Supplementary Parameters Theories.

## *7.2. Sensitive situations are rare*

Quantum Mechanics has been upheld in such a great variety of experiments, that Bell's Theorem might just appear as a proof of the impossibility of supplementary parameters. However, situations in which the conflict revealed by Bell's inequalities arises (sensitive situations) are so rare that, in 1965, none had been realized.

     To better appreciate this point, let us first note that Bell's inequalities are compatible with the whole classical physics, namely classical (relativistic) mechanics and classical electrodynamics, which can be imbedded into the Supplementary Parameters formalism obeying Einstein's causality. For instance, in classical mechanics, the $\lambda$'s would be the initial positions and velocities of the particles, from which the future evolution can be derived. Similarly, in classical electrodynamics, the $\lambda$'s would be the trajectories of the charges in the sources, from which one can deduce the electromagnetic fields, and their action on the measuring apparatus.

     Moreover, in situations usually described by quantum mechanics, it does not often happen that there is a conflict with Bell's inequalities. More precisely, for situations in which one looks for correlations between two separated subsystems (that may have interacted in the past), we can point out two conditions necessary to have the possibility of a conflict with Bell's inequalities :

- The two separated subsystems must be in an *entangled state,* non-factorizable, such as (1) (or the singlet state for two spin $1/2$ particles).

- For each subsystem, it must be possible to choose the measured quantity among at least two non-commuting observables (such as polarization measurements along directions **a** and **a'**, neither parallel nor perpendicular).

Even in such cases, we have seen that the conflict exists only for well chosen measured quantities (sets of orientations). But, as shown on figure 5, there are many orientations sets for which the quantum mechanical predictions do not violate Bell's inequalities.

It was realized in 1965 that there was no experimental evidence of a violation of Bell's inequalities. Since these inequalities are derived from very reasonable hypothesis, one could consider the possibility that the violation of Bell's inequalities indicate a situation where quantum mechanics fails. It was thus tempting to design a *sensitive experiment*, *i.e.* an experiment where the predictions of quantum mechanics for the real situation definitely violate Bell's inequalities. The experiment would then give a clearcut result between Quantum Mechanics, and Supplementary Parameter Theories obeying Bell's locality condition.

### *7.3. Production of pairs of photons in an EPR state*

As pointed out by C.H.S.H.[8], pairs of photons emitted in suitable atomic radiative cascades are good candidates for a sensitive test. Consider for instance a $J=0 \rightarrow J=1 \rightarrow J=0$ atomic cascade (Figure 6). Suppose that we select, with the use of wavelengths filters and collimators, two plane waves of frequencies $\nu_1$ and $\nu_2$ propagating in opposite directions along the $z$ axis (Figure 7).

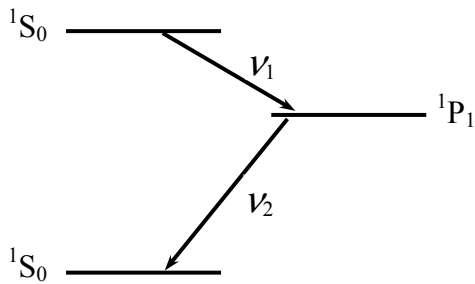
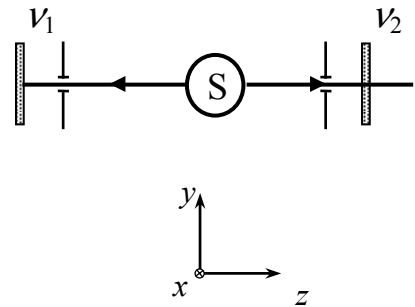

*Figure 6 - Radiative cascade emitting pairs of photons correlated in polarization.*

*Figure 7 - Ideal configuration (infinitely small solid angles).*

It is easy to show, by invoking parity and angular momentum conservation, that the polarization part of the state vector describing the pair $(\nu_1, \nu_2)$ can be written :

$$\Psi(\nu_1, \nu_2) = \frac{1}{\sqrt{2}}\left[|R,R\rangle + |L,L\rangle\right] \tag{29}$$

where $|R\rangle$ and $|L\rangle$ are circularly polarized states. By expressing $|R\rangle$ and $|L\rangle$ on a linear polarization basis, we obtain the state (1)

$$|\Psi(\nu_1,\nu_2)\rangle = \frac{1}{\sqrt{2}}\{|x,x\rangle + |y,y\rangle\}$$

With this entangled EPR state, one can envisage a sensitive experiment.

### *7.4. Realistic experiment*

A real experiment differs from the ideal one in several respects. For instance, the light should be collected in finite angles $2u$, as large as possible (Figure 8). In this situation, one can show[15] that the contrast of the correlation function decreases, since (6) is replaced by :

$$E_{QM}(\mathbf{a},\mathbf{b}) = F(u).\cos 2(\mathbf{a},\mathbf{b}) \qquad (30)$$

where $F(u) \leq 1$. Figure 9 displays $F(u)$ for a $J=0 \to J=1 \to J=0$ cascade. Fortunately, one can use large angles without great harm. For $u = 32°$ (our experiments), one has $F(u) = 0.984$.

All experimental inefficiencies (polarizers defects, accidental birefringences etc...) will similarly lead to a decrease of the correlation function $E(\mathbf{a},\mathbf{b})$. The function $S_{MQ}(\theta)$ (Figure 5) is then multiplied by a factor less than 1, and the conflict with Bell's Inequalities decreases, and even may vanish. Therefore, an actual experiment must be carefully designed and every auxiliary effect must be evaluated. All relevant parameters must be perfectly controlled, since a forgotten effect could similarly lead to a decrease of the conflict. For instance, an hyperfine structure dramatically decreases $F(u)$, so that only even isotopes should be used[15].

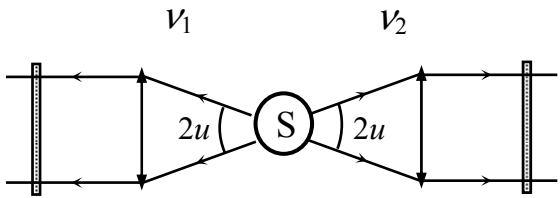
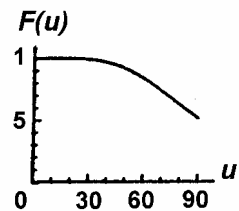

*Figure 8* - *Realistic configuration, with finite solide angles.*

*Figure 9* - *Reduction factor F(u) for a $J=0 \to J=1 \to J=0$ cascade*

### *7.5. Timing conditions*

As we have seen in section 6, Bell's Locality Condition may be considered a consequence of Einstein's Causality, if the experiment fulfils requirements, that can be split in two conditions :

i. The distant *measurements* on the two subsystems must be *space-like separated*.

ii. The *choices* of the quantities measured on the two separated subsystems must be made at *random*, and must be *space-like separated*.

The second condition is obviously more difficult to fulfil.

# 8. FIRST GENERATION EXPERIMENTS

The C.H.S.H. paper[8], published in 1969, had shown the possibility of realistic sensitive experiments with correlated photons produced in certain atomic cascades. Two groups started an experiment, one in Berkeley, one in Harvard. After their conflicting results, a third experiment was carried out in College Station (Texas). All the three experiments used a simplified experimental scheme, somewhat different from the ideal one since it involved one-channel polarizers.

## *8.1. Experiments with one channel polarizer*

In this simplified experimental scheme, one uses polarizers that transmit light polarized parallel to **a** (or **b**), but blocks the orthogonal one. Compared to the scheme of Figure 1, one can thus only detect the + results, and the coincidence measurements only yield the coincidence rates $N_{++}(\mathbf{a},\mathbf{b})$ between the + channels. In order to recover the missing − data, auxiliary runs are performed with one or both polarizers removed (the « orientation » of a removed polarizer is conventionally denoted ∞). We can then write relations between the *measured* coincidence rates $N_{++}(\mathbf{a},\mathbf{b})$, $N_{++}(\mathbf{a},\infty)$, and $N_{++}(\infty,\mathbf{b})$ and coincidence rates which are not measured:

$$N_{++}(\infty,\infty) = N_{++}(\mathbf{a},\mathbf{b}) + N_{+-}(\mathbf{a},\mathbf{b}) + N_{-+}(\mathbf{a},\mathbf{b}) + N_{--}(\mathbf{a},\mathbf{b})$$

$$N_{++}(\mathbf{a},\infty) = N_{++}(\mathbf{a},\mathbf{b}) + N_{+-}(\mathbf{a},\mathbf{b}) \qquad (31)$$

$$N_{++}(\infty,\mathbf{b}) = N_{++}(\mathbf{a},\mathbf{b}) + N_{-+}(\mathbf{a},\mathbf{b})$$

By substitution into the expression (28) of the polarisation correlation coefficient, and into inequalities (21), one can eliminate all the quantities which are not measured, and obtain new B.C.H.S.H. inequalities

$$-1 \le S' \le 0 \qquad (32)$$

where the quantity *S'*

$$S' = \frac{N(\mathbf{a},\mathbf{b}) - N(\mathbf{a},\mathbf{b'}) + N(\mathbf{a'},\mathbf{b}) + N(\mathbf{a'},\mathbf{b'}) - N(\mathbf{a'},\infty) - N(\infty,\mathbf{b})}{N(\infty,\infty)} \qquad (33)$$

is expressed as a function of *measured* coincidence rates only (we have omitted the implicit subscripts ++ in the expression above).

For the orientation sets shown on Figure 4, the Quantum Mechanical predictions violate the Bell's inequalities (32) :

$$S'^{Max}_{QM} = \frac{\sqrt{2}-1}{2} \qquad \text{for} \quad \theta = \pi/8$$

$$S'^{Min}_{QM} = \frac{-\sqrt{2}-1}{2} \qquad \text{for} \quad \theta = 3\pi/8 \qquad (34)$$

It is therefore possible to make a sensitive test with one channel polarizers also.

Note however that the derivation of the Bell's inequalities (32) requires a *supplementary assumption*. Since the detection efficiencies are low (due to small collection angle and low photomultipliers efficiencies), the probabilities involved in the expression of $E(\mathbf{a},\mathbf{b})$ must be redefined on the ensemble of pairs that would be detected with polarizers removed. This procedure is valid only if one assumes a reasonable hypothesis about the detectors. The C.H.S.H. assumption[8] states that, « given that a pair of photons emerges from the polarizers, the probability of their joint detection is independent of the polarizer orientations » (or of their removal). Clauser and Horne[11] have exhibited another assumption, leading to the same inequalities. The status of these assumptions has been thoroughly discussed in reference 16.

## *8.2. Results*

In the Berkeley experiment[17], Clauser and Freedman built a source where calcium atoms were excited to highly lying states by ultraviolet radiation. The atom would then decay, and among the various desexcitation routes, it had some probability to emit a couple of green and violet correlated photons ($4p^2\ {}^1S_0 \to 4s4p\ {}^1P_1 \to 4s^2\ {}^1S_0$ radiative cascade). Since the signal was weak, and spurious cascades occurred, it took more than 200 hours of measurement for a significant result. The results were found in agreement with Quantum Mechanics, and a violation of the relevant Bell's inequalities (32) was observed (by 5 standard deviations).

At the same time, in Harvard, Holt and Pipkin[18] found a result in disagreement with Quantum Mechanics, and in agreement with Bell's Inequalities. Their source was based on the $9^1P_1 \to 7^3P_1 \to 6^3P_0$ cascade of Mercury (isotope 200), excited by electron bombardment. The data accumulation lasted 150 hours. Clauser subsequently repeated their experiment, but with Mercury 202. He found an agreement with Quantum Mechanics, and a significant violation of Bell's Inequalities[19].

In 1976, in Houston, Fry and Thompson[20] built a much improved source of correlated photons, emitted in the $7^3S_1 \to 6^3P_1 \to 6^3S_0$ cascade in Mercury 200. This is a $J=1 \to J=1 \to J=0$ cascade, a priori not as favorable as a $J=0 \to J=1 \to J=0$ cascade, but they could selectively excite the upper level of the cascade, by use of a *C.W.* single line tunable laser (a quite rare instrument at that time). The signal was several order of magnitude larger than in previous experiments, allowing them to collect the relevant data in a period of 80 minutes. Their result was in excellent agreement with Quantum Mechanics, and they found a violation, by 4 standard deviations, of the Bell's inequalities (32) specific of single channel polarizers experiments.

# 9. ORSAY EXPERIMENTS (1980-1982)[14]

## *9.1. The source*

From the beginning of our programme, our goal was to implement more sophisticated experimental schemes[13], so we devoted a lot of efforts to develop a high-efficiency, stable, and well controlled *source of entangled photons*. This was achieved (Figure 10) by a two photon selective excitation[21] of the $4p^2\ {}^1S_0 \to 4s4p\ {}^1P_1 \to 4s^2\ {}^1S_0$ cascade of calcium already used by Clauser and Freedman. This cascade is very well suited to coincidence experiments, since the lifetime $\tau_r$ of the intermediate level is rather short (5 *ns*). If one can reach an excitation rate of about $1/\tau_r$, then an optimum signal-to-noise ratio for coincidence measurements with this cascade is reached.

We were able to obtain this optimum rate with the use of a Krypton ion laser ($\lambda_K = 406\,nm$) and a tunable dye laser ($\lambda_D = 581\,nm$) tuned to resonance for the two-photon process. Both lasers were single-mode operated. They were focused onto a Calcium atomic beam (laser beam waists about 50 *µm*). Two feedback loops provided the required stability of the source (better than 0.5 % for several hours): the first loop controlled the wavelength of the tunable laser to ensure the maximum fluorescence signal; a second loop controlled the angle between the lasers polarisations, and compensated all the fluctuations of the cascade rate. With a few tens of milliwatts from each laser, the cascade rate was about $N = 4 \times 10^7 s^{-1}$. An increase beyond this rate would not have significantly improved the signal-to-noise ratio for coincidence counting, since the accidental coincidence rate increases as $N^2$, while the true coincidence rate increases as *N*. At this cascade rate, the coincidence rate with parallel polarizers was about $10^2\ s^{-1}$, several orders of magnitude larger than in the first experiments. A statistical accuracy of 1% could then be achieved in each individual run of duration 100 *s*.

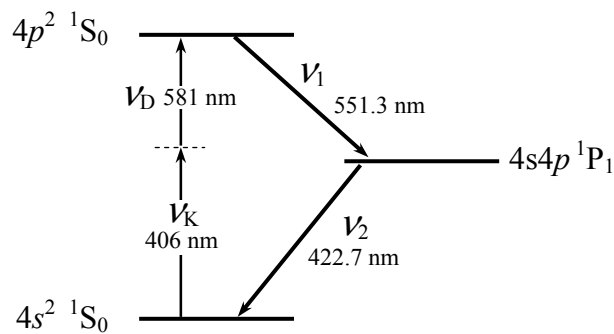

*Figure 10* - *Two-photon selective excitation of the 4p$^2$ ${}^1S_0$ state of Calcium with a Krypton ion laser and a tunable dye laser. From this state, the atom radiative decay can only deliver the pair of entangled photons ($v_1, v_2$).*

## 9.2. Detection - Coincidence counting

The fluorescence light was collected by two large-aperture aspherical lenses ($u = 32°$, as defined on figure 8), followed in each leg by an interference filter (respectively at 551.3 *nm* and 422.7 *nm*), a transport optical system, a polarizer, and a photomuliplier tube. The photomultipliers fed the coincidence-counting electronics, that included a time-to-amplitude converter and a multichannel analyzer, yielding the time-delay spectrum of the two-photon detections (Figure 11). This spectrum first shows a flat background due to accidental coincidences (between photons emitted by different atoms). The true coincidences (between photons emitted by the same atom) are displayed in the peak rising at the null-delay, and exponentially decaying with a time constant $\tau_r = 5$ *ns* (lifetime of the intermediate state of the cascade). The measured coincidence signal is thus the area of the peak.

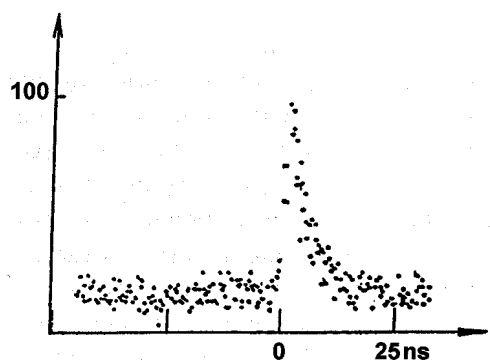

*Figure 11 - **Time-delay spectrum**. Number of detected pairs as a function of the delay between the detections of two photons. The flat background corresponds to accidental coincidences between uncorrelated photons emitted by different atoms, and scales as the square $N^2$ of the cascade rate. The peak, whose area scales as N, corresponds to correlated photons, and gives the coincidence rate to be measured.*

Additionally, a standard coincidence circuit with a 19 *ns* coincidence window monitored the rate of coincidences around null delay, while a delayed-coincidence channel monitored the accidental rate. It was then possible to check that the true coincidence rate obtained by substraction was equal to the signal in the peak of the time-delay spectrum.

In the second and third experiments, we used a fourfold coincidence system, involving a fourfold multichannel analyzer and four double-coincidence circuits. The data were automatically gathered and processed by a computer.

## 9.3. Experiment with one-channel polarizers[22]

Our first experiment was carried out using one channel pile of plates polarizers, made of ten optical grade glass plates at Brewster angle, ensuring an excellent rotational invariance. For a fully polarized light, the maximum and minimum transmission were $0.975 \pm 0.005$ and $0.030 \pm 0.005$ respectively.

Thanks to our high-efficiency source, allowing us to achieve an excellent statistical accuracy in 100 s runs, we could perform various statistical checks, as well as physical checks, for instance on the rotational invariance of the signals (for all these measurements, the long term stability of the source, at the level of 0.5%, was found crucial).

A direct test of the Bell's inequalities for single channel polarizers (32) has been performed. We have found for the quantity *S'* (equation 33)

$$S'_{exp} = 0.126 \pm 0.014 \qquad (35)$$

violating inequalities (32) by 9 standard deviations, and in good agreement with the Quantum Mechanical predictions for our polarizers efficiencies and lenses aperture angles :

$$S'_{QM} = 0.118 \pm 0.005 \qquad (36)$$

The uncertainty in the theoretical value $S'_{QM}$ accounts for the uncertainty in the measurements of the polarizers efficiencies.

The agreement between the experimental data and the Quantum Mechanical predictions has been checked in a full 360 ° range of orientations (Figure 12).

We have repeated these measurements with the polarizers moved at 6.5 m from the source. At such a distance (four coherence-lengths of the wave packet associated with the lifetime $\tau_r$) the *detection events were space-like separated*, and we therefore fulfilled the first time condition of section 7.5. No modification of the experimental results was observed, and the Bell's inequality was violated by the same amount.

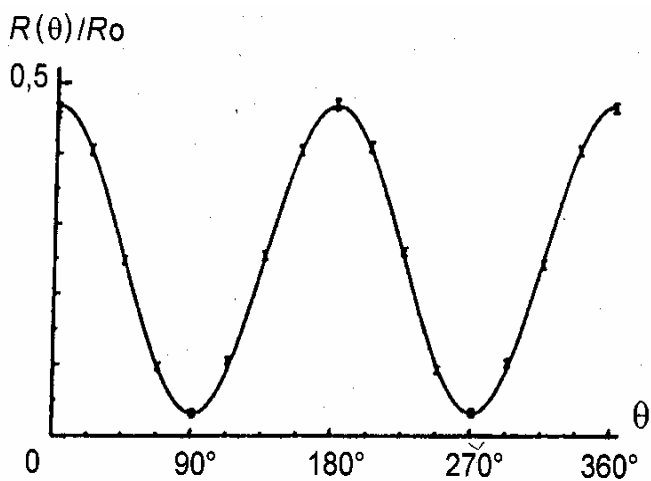

*Figure 12 - **Experiment with one channel polarizers.** Normalized coincidence rate as a function of the relative polarizers orientation. Indicated errors ± 1 standard deviation. The solid curve is not a fit to the data but the prediction by Quantum Mechanics for the actual experiment.*

### 9.4. Experiment with two-channel analyzers[23,24,25]

With single-channel polarizers, the measurements of polarization are inherently incomplete. When a pair has been emitted, if no count is obtained at one of the photomultipliers, there is no way to know if « it has been missed » by the detector or if it has been blocked by the polarizer (only the later case corresponds to a − result for the measurement). This is why one had to resort to auxiliary experiments, and indirect reasoning using supplementary assumptions, in order to test Bell's inequalities.

With the use of two-channel polarizers, we have performed an experiment following much more closely the ideal scheme of Figure 1. Our polarizers were polarizing cubes with dielectric layers transmitting one polarization and reflecting the orthogonal one††. Such a polarization splitter, and the two corresponding photomultipliers, are fixed on a rotatable mount. This device (polarimeter) yields the + and − results for a linear polarization measurement. It is an optical analog of a Stern-Gerlach filter for a spin 1/2 particle.

With polarimeters *I* and *II* in orientations **a** and **b**, and a fourfold coincidence counting system, we are able to measure in a single run the four coincidence rates $N_{\pm\pm}(\mathbf{a},\mathbf{b})$, and to obtain directly the polarization correlation coefficient $E(\mathbf{a},\mathbf{b})$ by plugging the numbers into equation (28). It is then sufficient to repeat the same measurment for a sensitive set of four orientations, and the ideal Bell's inequality (20) can be directly tested.

This experimental scheme being much closer to the ideal scheme of figure 1 than previous experiments with one channel polarizers, we do not need the strong supplementary assumption on the detectors. However, the detection efficiency in each channel is well below unity, first because of the limited solid angle of collection, and second because of the efficiency of the photomultiplier. An advocate of hidden variable theories could then argue that we are not sure that the sample on which the measurement bears, remains the same when the orientations of the polarimeters are changed. In order to be logically allowed to compare our measurements to Bell's inequalities, we therefore also need a supplementary assumption: we must assume that the ensemble of actually detected pairs is independent of the orientations of the polarimeters. This assumption is very reasonable with our symmetrical scheme, where the two orthogonal output channels of a polarizer are treated in the same way (the detection efficiencies in both channels of a polarimeter are equal). Moreover, we have experimentally checked that the sum of the four coincidence rates $N_{\pm\pm}(\mathbf{a},\mathbf{b})$ remains constant when the orientations change, although each coincidence rate is 100% modulated. This shows that the size of the selected sample of pairs is constant. Of course, it is not a proof of the validity of the assumption, but at least it is consistent with it. Note that it is possible to use a stronger assumption, the «*fair sampling assumption* », in which one assumes that the ensemble of detected pais is a fair sample of the ensemble of all emitted pairs. The assumption above is a logical consequence of the fair sampling assumption, which is stronger. On the other hand, the fair sampling assumption is very reasonable, and easy to express.

The experiment has been done at the sensitive set of orientations of Figure 4a, for which a maximum conflict is predicted. We have found

---

†† A similar experiment, using calcite two channel polarizers, had been considered at the University of Catania[26].

$$S_{exp} = 2.697 \pm 0.015 \qquad (37)$$

violating the inequalities (21) $(|S| \leq 2)$ by more than 40 standard deviations! Note that this result is in excellent agreement with the predictions of Quantum Mechanics for our polarizers efficiencies and lenses apertures :

$$S_{QM} = 2.70 \pm 0.05 \qquad (38)$$

The uncertainty indicated for $S_{QM}$ accounts for a slight lack of symmetry of the two channels of a polarizer (± 1%). The effect of these dissymetries has been computed and cannot create a variation of $S_{QM}$ greater than 2 %.

We have also performed measurements of the polarization correlation coefficient $E(\mathbf{a},\mathbf{b})$ in various orientations, for a direct comparison with the predictions of Quantum Mechanics (Figure 13). The agreement is clearly excellent.

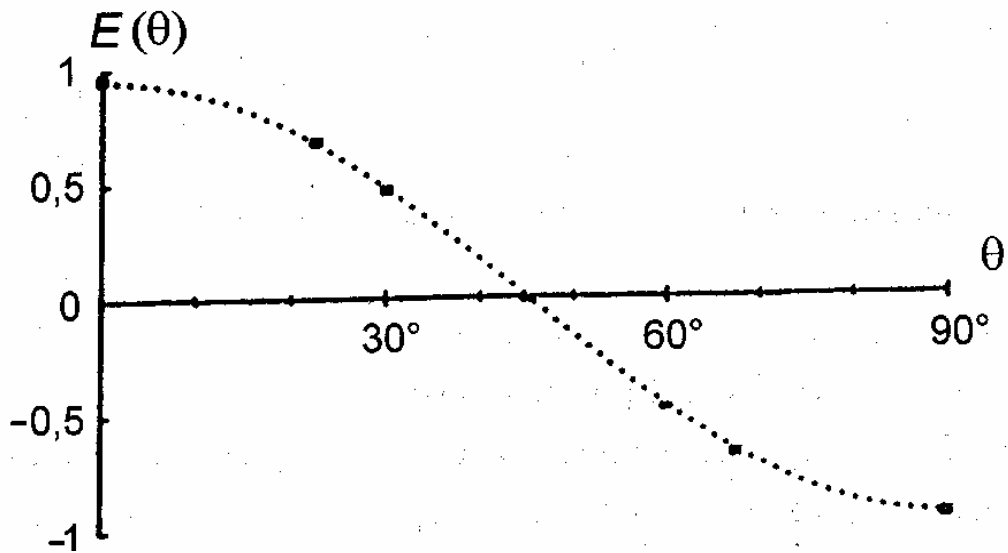

*Figure 13 - **Experiment with two-channels polarizers**. Polarization correlation of as a function of the relative angle of the polarimeters. The indicated errors are ± 2 standard deviations. The dashed curve is not a fit to the data, but Quantum Mechanical predictions for the actual experiment. For an ideal experiment, the curve would exactly reach the values ± 1.*

These measurements can be presented in a different way, to emphasize the relevance to the test of Bell's inequalities. On figure 14, we show the measured quantity $S(\theta)$, as defined in section 4.2. The violation of Bell's inequalities is clear around 22.5° (which corresponds to the result (37)) and 67.5°, but one also sees, as already emphasized, that there are many situations where there is no conflict with Bell's inequalities.

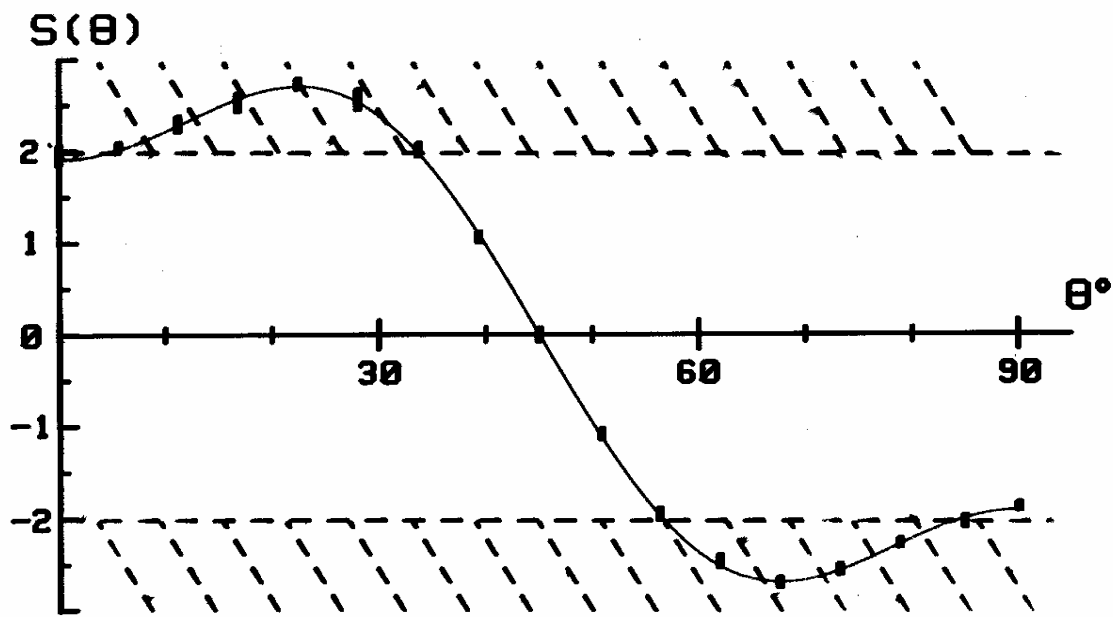

**Figure 14 - Experiment with two-channels polarizers**. Quantity S(θ), to be tested by Bell's inequalities (−2 ≤ S ≤ +2), as a function of the relative angle of the polarimeters. The indicated errors are ± 2 standard deviations. The dashed curve is not a fit to the data, but Quantum Mechanical predictions for the actual experiment. For an ideal experiment, the curve would exactly reach the values ± 2.828.

### *9.5. Timing experiment[27]*

As stressed in sections 6 and 7.5, an ideal test of Bell's inequalities would involve the possibility of switching at random times the orientation of each polarizer[13], since the locality condition would become a consequence of Einstein's causality. We have done a step towards such an ideal experiment by using the modified scheme shown on Figure 15.

In that scheme[13], each (single-channel) polarizer is replaced by a setup involving a switching device followed by two polarizers in two different orientations : **a** and **a'** on side *I*, **b** and **b'** on side *II*. The optical switch $C_1$ is able to rapidly redirect the incident light either to the polarizer in orientation **a**, or to the polarizer in orientation **a'**. This setup is thus equivalent to a variable polarizer switched between the two orientations **a** and **a'**. A similar set up is implemented on the other side, and is equivalent to a polarizer switched between the two orientations **b** and **b'**. In our experiment, the distance *L* between the two switches was 13 *m*, and *L / c* has a value of 43 *ns*.

The switching of the light was effected by home built devices, based on the acousto-optical interaction of the light with an ultrasonic standing wave in water. The incidence angle (Bragg angle) and the acoustic power, were adjusted for a complete switching between the 0$^{th}$ and 1$^{st}$ order of diffraction. The switching function was then of the form $\sin^2(\frac{\pi}{2}\cos\Omega_a t)$, with the acoustic frequency $\Omega_a / 2\pi$ of the order of 25 *MHz*.

The change of orientation of the equivalent variable polarizer then occurred after inequal intervals of 6.7 *ns* and 13.3 *ns*. Since these intervals as well as the delay between the emissions of the two photons of a pair (average value of $\tau_r = 5 ns$), were small compared to $L/c$ (43 *ns*), a detection event on one side and the corresponding change of orientation on the other side were separated by a space-like interval. The first time condition was clearly fulfilled. The second time-condition was almost fulfilled, except for the fact that the switching was not truly random, but rather quasiperiodic (we discuss this point below).

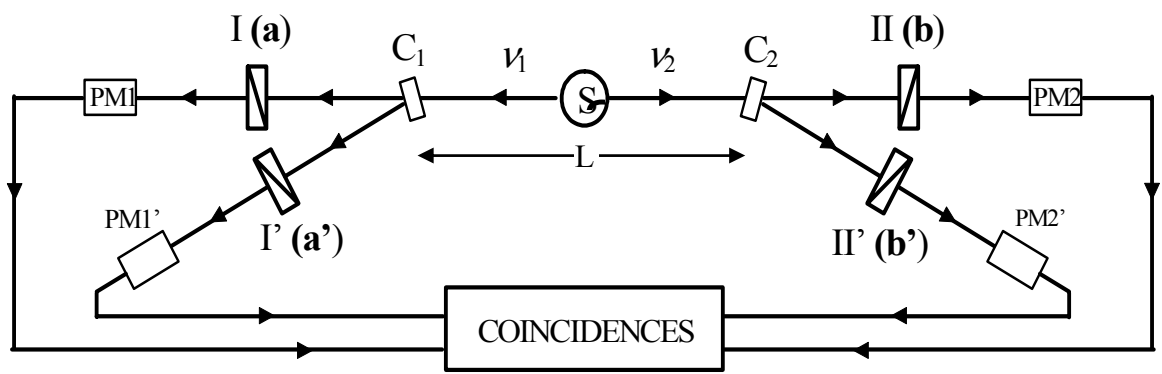

*Figure 15 - **Timing-experiment with optical switches** ($C_1$ and $C_2$). The switch $C_1$ followed by the two polarizers in orientations* **a** *and* **a'** *is equivalent to a single polarizer switched between the orientations* **a** *and* **a'**. *A switching occurs approximatively each* 10 *ns. A similar setup, independently driven, is implemented on the second side.In our experiment, the distance* L *between the switches was large enough* (13 m) *that the time of travel of a signal between the switches at the velocity of light* (43 ns) *was significantly larger than the delay between two switchings (about* 10 ns *) and the delay between the emission between the two photons* (5 ns *average).*

The experiment was far from ideal on other points. First, in order to match the photon beams to the aperture of the switches, we had to reduce their size by a factor of 3, entailing a reduction of the coincidence rates by one order of magnitude. As a consequence, to achieve a significant statistical accuracy, the duration of data accumulation was much longer than in previous experiments, and we had to face problems of drifts. It was then necessary to average out the various measured quantities. Second, for not infinitely small beams, the commutation by the switches is incomplete, because the incidence angle is not exactly the Bragg angle for all rays. In our experiment, the minimum of the light intensity in each channel was 20%, so that not all photons were submitted to forced switching. Third, in this experiment, we used single channel polarizers, which allowed us to do it with the same fourfold coincidence system as in the the static experiment of section 9.4.

Our test of Bell's Inequalities involved a total of 8000 *s* of data accumulation with the 4 polarizers in the orientations of Figure 4.a. A total of 16000 *s* was devoted to the measurements with half or all the polarizers removed. In order to compensate the effects of systematic drifts, the data accumulation was alternated between the various configurations each 400 *s*, and the data were averaged out. We finally obtained

$$S'_{\text{exp}} = 0.101 \pm 0.020 \qquad (39)$$

violating the upper limit of the Bell's inequality (32) by 5 standard deviations, and in good agreement with the Quantum Mechanics predictions

$$S'_{QM} = 0.113 \pm 0.005 \tag{40}$$

Other measurements of the coincidence rates were carried out, for a comparison with Quantum Mechanics at different angles. As shown on Figure 16, the results were in good agreement with the predictions of Quantum Mechanics.

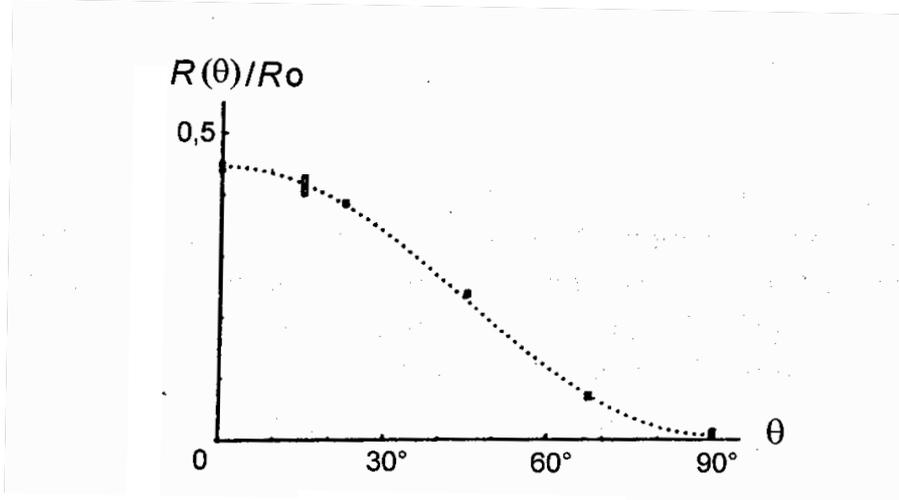

*Figure 16 - Timing experiment: average normalized coincidence rate as a function of the relative orientation of the polarizers. Indicated errors are ± 1 standard deviation. The dashed curve is not a fit to the data but the predictions by Quantum Mechanics for the actual experiment.*

According to these results, Supplementary-Parameters Theories obeying Einstein's Causality seem to be untenable. However, as indicated earlier, our experiment was not ideal, from several points of view, and several loopholes were left open for a strict advocate of hidden variable theories. First, because we used single channel polarizers, the experiment is significant only if one accepts a strong version of the « fair sampling » assumption. Adressing more specifically the timing aspect of this experiment, an advocate of hidden variable theories might argue that the switching was not complete, and that only the pairs undergoing forced switching must obey Bell's inequalities. But since these pairs represent a large fraction of the total number, it is hard to believe that we would not have observed a significant discrepancy between our results and the Quantum Mechanical predictions.

The most important point to discuss is the fact that the switches were not truly random, since the acousto-optical devices were driven by periodic generators. Note however that the two generators on the two sides were functionning in a completely uncorrelated way, since they were operated by different RF generators at different frequencies (23.1 *MHz* and 24.2 *MHz*), with uncorrelated frequency drifts. Moreover, another random feature is provided by the random delay between the two photons of a pair (exponentially decaying distribution of constant $\tau_r = 5ns$, as shown on figure 11) which are distributed on an interval larger than the time between two successive switchings.

In conclusion, this experiment, which was until 1998 the only one involving forced fast changes of the settings of the analysers, had enough imperfections to leave open the possibility of ad hoc supplementary parameter models fulfilling Einstein's causality. However, several models that we have tried are eliminated by our experimental results, which are constituted not only by the measured value (39) of *S'*, but also by the time delays spectra without any accident observable on the exponential decay, and with areas in good agreement with quantum mechanics as shown on figure 15. We think that these data should be taken into account by any advocate of local hidden variable theories trying to build a model compatible with experimental observations.

## 10. THIRD GENERATION : EXPERIMENTS WITH PAIRS OF PHOTONS PRODUCED IN PARAMETRIC DOWN CONVERSION

As we have already noted, the calcium radiative cascade that was used in our experiments was excited to an optimum rate, beyond which there is not much possibility of gain in signal to noise ratio. Since the life time of the intermediate stage is quite short (5 ns) the situation is very favourable for coincidence counting, and there was not much room left for improvement with sources based on atomic radiative cascades[28]:

In the late 80's, new sources of pairs of correlated photons have been developed simultaneously by two groups[29,30]. In these sources, a pair of red photons is produced by parametric down conversion of a U.V. photon. Because of the phase matching condition in the non linear crystal used for this process, there is a strong correlation between the directions of emission of the two photons of a pair, so that, by spatial selection with two diaphragms positionned in conjugate positions, one can in principle be sure to get the two photons of a pair. This is in contrast with the atomic radiative cascades which produce photons only weakly correlated in direction[15] : since each photon is collected in a solid angle $\Omega$ small compared to $4\pi$, the probability to get the second photon of a pair, once a first one is detected, is of the order of $\Omega/4\pi$ , so that the sample of detected pairs is smaller by this factor, than the sample of selected pairs. The new scheme with photons correlated in direction allows one to get rid of this reduction factor, and this has far reaching consequences, both practical and fundamental. On the practical side, it allows larger coincidence rates to be obtained, for similar cascade rates: in the most favourable case[31] the coincidence rate may be more than one order of magnitude larger than in our best experiments (section 9). Moreover, such large coincidence rates can be obtained with narrow photon beams (with a small Fresnel number). Such beams can easily be matched into small optical components, or even optical fibers, which opens many new possibilities.

These new sources can produce pairs or photons correlated in polarisation[29,30,31,32], in states analogous to (1). But they can also produce entangled states exhibiting EPR type correlations between observables other than polarization. An interesting case[33] considers pairs of photons where each photon is emitted « at two different times ». Here, the relevant observable is the time of emission of the two photons of the pair, and the conjugate one is the energy (wavelength). Corresponding experiments have been carried out[34,35,36]. Note that this scheme, where polarization is not the relevant observable, is specially interesting for experiments with optical fibers, in which polarization control may be a crucial issue. Another interesting scheme considers the directions of emission as observables[37] : each photon of an entangled pair involves two different directions of emissions, strongly correlated to two directions of emission for the second photon. An experiment of this type has also been carried out[38].

As emphasised in reference [37], all these new schemes can be embedded in the general framework of « two particles interferences »: indeed, the joint measurements probabilities are the square of a sum of two amplitudes (each involving the two photons), with a relative phase that can be controlled experimentally. Although it was not pointed out by the authors of ref. [37], the original EPRB scheme is a very clear example of this situation. For instance, for the polarization entangled state of section 2, the state (1) can also be rewriten (see equation 29) as the superposition of a state $|L,L\rangle$ where both photons have a left handed helicity, and a state $|R,R\rangle$ with two right handed helicities. For each of these two states, the amplitude for being detected in any couple of output channels behind the linear polarizers (see figure 1) has a value of $1/\sqrt{2}$ times a phase factor which depends on the orientation of the polarizers (Figure 1). The addition of the amplitudes associated to $|L,L\rangle$ and $|R,R\rangle$ thus leads to an interference term responsible for the sinusoidal variations of the joint probabilities (3) when the orientations change[14].

These new sources and schemes have lead to a series of tests of Bell's inequalities, *which have all confirmed quantum mechanics*. Clear violations of Bell's inequalities have been found, under the assumption that the « fair sampling hypothesis » holds. Among these, it is worth pointing out a violation of Bell's inequalities by 100 standard deviations in a few minuts only[31]. Note also an experiment[35] where a clear violation of Bell's inequalities has been observed with one leg of the apparatus made of 4 kilometers of optical fiber. More recently, EPR correlations have been observed with photons propagating in several tens of kilometers of commercial telecommunication fibers[39].

These experiments of third generation should ultimately lead to an ultimate experiment where there would be no remaining loophole left open. First, the perfect correlation between the directions of emission offers the possibility to close the loophole related to the low detection efficiency[11], when photon detectors with quantum efficiency close to unity are available[40].

The second class of fundamental improvements is related to the « timing experiments » (section 7.5, and 9.5). Ideally[13], one needs polarizers that can be independently reorientated at random times, with a reorientation autocorrelation time shorter than the space separation $L/c$ between the polarizers. Our third experiment (section 9.5), which was the first attempt in this direction, was basically limited by the wide size of the beams carrying the correlated photons: this prevented us to use small size electrooptic devices suitable for random switching of polarization. With the new schemes using optical fibers, it becomes possible to work with small integrated electrooptical devices.Moreover, with use of optical fibers, the detectors can be kilometers apart. At such separations (several microseconds), the time conditions become less stringent, and truly random active operation of the polarizers become possible at this time scale[41]. An experiment of this type has been completed in the group of Anton Zeilinger[42]. From the point of view of the timing condition, one can say that this experiment meets all the criteria of an ideal experiment[43].

## 11. CONCLUSION

We have nowadays an impressive amount of sensitive experiments where Bell's inequalities have been clearly violated. Moreover, the results are in excellent agreement with the quantum mechanical predictions including all the known features of the real experiment. Each of the remaining loophole has been separately closed [40,42], and although yet more ideal experiments are still desirable [44], it is legitimate to discuss the consequences of the rejection of supplementary parameter theories obeying Einstein's causality.

It may be concluded that quantum mechanics has some non-locality in it, and that this non-local caracter is vindicated by experiments [45]. It is very important however to note that such a non-locality has a very subtle nature, and in particular that it cannot be used for faster than light telegraphy. It is indeed simple to show [46] that in a scheme where one tries to use EPR correlations to send a message, it is necessary to send a complementary information (about the orientation of a polarizer) via a normal channel, which of course does not violate causality. This is similar to the teleportation schemes [47] where a quantum state can be teleported via a non-local process, provided that one also transmits classical information via a classical channel. In fact, there is certainly a lot to understand about the exact nature of non-locality, by a careful analysis of such schemes [48].

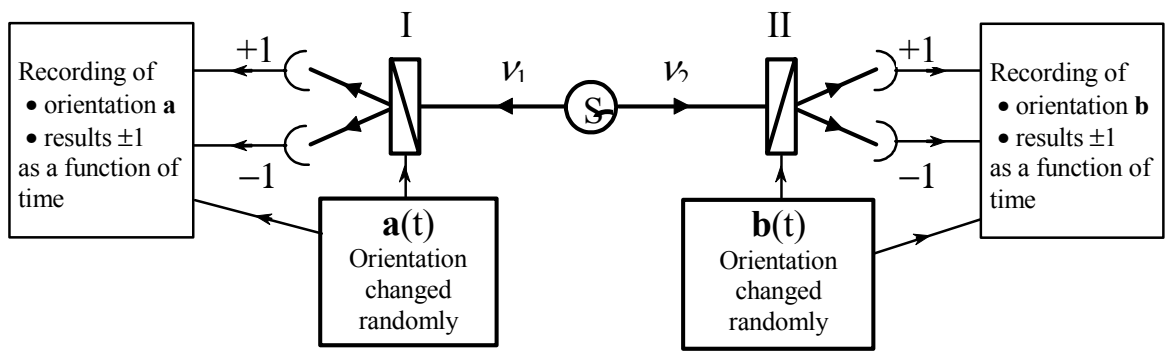

*Figure 17. Ideal timing experiment.* Each polarizer is randomly reoriented during the propagation of photons between the source and the polarizers. On each side, one records the orientation of the polarizer as well as the results of polarization measurements as a function of time. When a run is completed, the two data sets from the two sides are brought together, and one can determine the value of the correlation as a function of the relative orientation at the moment of the measurement.

When realizing that this quantum non locality does not allow one to send any useful information, one might be tempted to conclude that in fact there is no real problem, and that all these discussions and experimental efforts are pointless. Before rushing to this conclusion, I would suggest to consider an ideal experiment done with the scheme of figure 17. On each side of the experiment of Fig. 1, done with variable analysers, there is a monitoring system, that registers the detection events in channels + or − with their exact dates. We also suppose that the orientation of each polarizer is changed at random times, also monitored by the system of the corresponding side. It is only when the experiment is completed that the two sets of data, separately collected on each side, are brought together, in order to extract the correlations. Then, looking into the data that were collected previously, and that correspond to paired events that were space like separated when they happened, one can see that indeed the correlation did change at the very moment when the relative orientation of the polarizers changed.

So when one takes the point of view of a delocalized observer, which is certainly not inconsistent when looking into the past, it must be acknowledged that there is a non local behaviour, in the EPR correlations. Entanglement is definitely a feature going beyond any spacetime description *à la* Einstein: a pair of entangled photons must be considered a single global object, that we cannot consider as made of individual objects separated in spacetime with well defined properties.

For many years, I have been quoting the scheme of Figure 17 as a GedankenExperiment useful for the sake of the discussion. Nowadays, we are lucky that this experiment has been done in the real world: the experiment of Zeilinger and Weihs[42], sketched on Figure 18, exactly follows[43] the scheme of Figure 17. Once again, the EPR problem has switched from fundamental discussions bearing on GedankenExperiment, to real experiments. *We must be grateful to John Bell for having shown us that philosophical questions about the nature of reality could be translated into a problem for physicists, where naive experimentalists can contribute.*

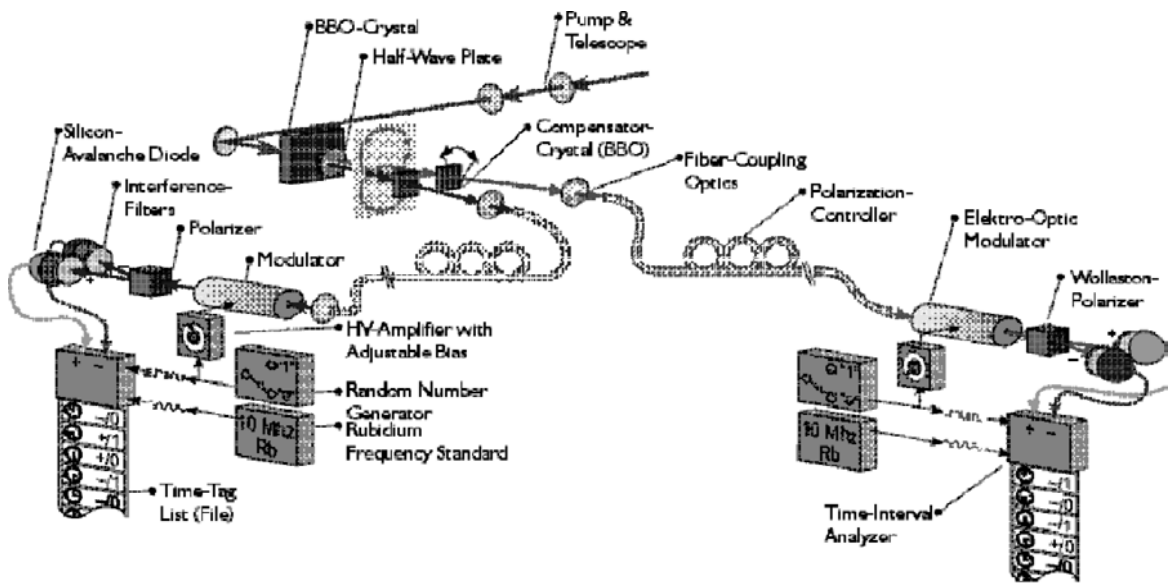

*Fig. 18. The timing experiment of Weihs et al. This experiment follows closely the ideal scheme of Figure 17, since the two ends of the experiment are totally independent. It is only after completion of a run that the data from the two sides are compared, in order to extract the correlation coefficient. Note also that the orientations of the polarizers are changed randomly during the photon propagation from the source to the polarizers.*

---